# Representative longitudinal optical phonon modes in polar semiconductor quantum dots

Tiberius O. Cheche[*a], Valentin Barna[a], Ioan Stamatin[a]

[a]University of Bucharest, Faculty of Physics, Bucharest 077125, Romania

**Abstract**

Existence of representative longitudinal optical (LO) phonon modes is theoretically discussed for the case of polar semiconductor cylindrical quantum dots embedded in a semiconductor matrix. The approach is developed within the dielectric continuum model considering the Fröhlich interaction between electrons and the confined LO phonons. The theory is applied to cylindrical GaAs/AlAs quantum dots within an adiabatic treatment.

## 1. Introduction

In terms of Born-Oppenheimer approach the optical transition is a process by which the electron jumps between different electronic potential energy surfaces with emission or absorption of photons. The potential energy surfaces are functions of the configuration nuclear coordinates, the nuclei displacements relative to equilibrium position. The vibrational modes of the system are obtained from the nuclear Schrödinger equation of the Born-Oppenheimer approach. Reduction of the number of vibrational modes involved in the transfer to the *representative* mode(s) by introducing the so-called reaction coordinate(s) is a well-known tool for the theoretical treatment of the electron transfer in chemical physics processes [1]. In semiconductors nanostructures use of the theoretical treatments based on the picture of representative phonon modes interacting with the (multi)excitonic states is less common. Introduction of such representative phonon mode(s) is desirable for a theoretical investigation of the relaxation processes when the exciton-phonon interaction is the key factor in the analyzed phenomenon. Due to the polar coupling to longitudinal optical (LO) phonons, the carriers lose their initial kinetic energy to the lattice by a very efficient relaxation mechanism in III-V bulk semiconductors or quantum wells. Experiments proved a similar efficiency of relaxation by LO phonons in the case of semiconductor quantum dots (QDs) [2].

Adiabatic [3] and non-adiabatic [4] approaches are used in investigation of the optical processes and relaxation in QDs. When the electron oscillation is much faster than that of the heavy ions (strong geometrical confinement or strong electron-phonon coupling) an adiabatic treatment is appropriate. On the other hand, presence of the dynamic Jahn-Teller effect (existence of degenerate electronic levels) or the pseudo-Jahn-Teller effect (phonon and the inter-level electronic energies are of close values) may impose a non-adiabatic treatment.

There is rich literature on the exciton-LO phonon interaction in nanostructures of different shapes, size, and chemical composition. In this work we consider the problem of cylindrical QDs [5-10] and based on an analytic model we obtain the exciton-LO phonon interaction in QDs of cylindrical shape. We adopt a simple analytical solvable QD model within the adiabatic approach and consider a dielectric continuum treatment for the confined phonons. Within this framework, we prove existence of the representative confined bulk-like LO phonon modes, stronger coupled to the exciton. The paper is organized as follows. In section 2 the analytically solvable model for semiconductor cylindrical QDs is introduced. We consider small QDs and consequently, in first approximation, we neglected the Coulombic electron-hole interaction. According to this simplification we approximate the exciton states by electron-hole pairs (EHPs). In section 3, the confined bulk-like LO phonons are obtained for polar semiconductors by a dielectric continuum approach and expression of the electron-LO phonon interaction is found considering Fröhlich electron-phonon coupling. In section 4, the theory is applied to GaAs/AlAs cylindrical QDs, the Huang-Rhys factor is evaluated, and the representative LO phonon mode is identified. The last section contains conclusions.

## 2. Analytic solvable quantum dot model

In the effective mass approximation the electron single particle wave function of QD may be approximated as $\Psi_{\alpha\sigma}(\mathbf{r}) = u_\sigma(\mathbf{r})\psi_\alpha(\mathbf{r})$, where $\mathbf{r}$ is the position vector of the carrier. $\psi_\alpha^e(\mathbf{r})$ is the envelope wavefunction, and $u_\sigma(\mathbf{r})$ is the periodical Bloch function at the $\Gamma$ point with spin dependence included. The same is valid for holes by replacing $e$ by $h$, $\alpha$ by $\beta$, and $\sigma$ by $\tau$. $\sigma$ and $\tau$ are the $z$-projections of the Bloch angular momentum, with $\sigma = \pm 1/2$ and $\tau = \pm 3/2, \pm 1/2$. In GaAs QDs, the type of QD to which the theory is applied in this work, the degeneracy of the heavy hole and light hole band is lifted due to the confinement. For the lowest lying hole states the expected predominantly heavy hole character is confirmed experimentally [11]. Consequently, by disregarding the band-mixing, we assume that the topmost states are formed from degenerate heavy-hole states and within a two-band approach we have $\tau = \pm 3/2$.

We consider QD of cylindrical shape with radius $R$, and height $2d$. The effective masses that are considered isotropic are denoted by $m_e$, and $m_h$, for electrons, and holes, respectively. We use the cylindrical coordinates $\rho, z, \varphi$. The confinement potentials in $z$ direction are established by the band-offset $V_b^{e,h}$ for electron, holes: $V(z) = 0$ for $|z| < d$, and $V(z) = V_b^*$ for $|z| \geq d$, with $V_b^*$ standing for the electron or hole potential. In radial direction the potential is taken infinite for $\rho \geq R$. For such a confinement potential the Schrödinger equation of the single particle envelope wave function is separable. Next, for convenience, we give up to the index representing the type of carrier. Then, i) by introducing the energy dispersion relation for the energy structure of the QD, $E(n,p) = \varepsilon(n) + \hbar^2 p^2/(2m_*)$, with $m_*$ the generic effective mass of the carrier (electron or hole), and $p$ and $n$ parameters to be determined, and ii) by choosing envelope wave function of form $\psi(\mathbf{r}) = C J(p\rho) f(nz) \exp(im\varphi)$ with $C$ normalization constant, we obtain the simple equations:

$$\frac{-\hbar^2}{2m_*(z)} \frac{\partial^2 f(nz)}{\partial z^2} + V(z)f(nz) = E(n)f(nz) \quad (1a)$$

$$\frac{1}{\rho}\frac{\partial}{\partial \rho}\left(\rho \frac{\partial J(p\rho)}{\partial \rho}\right) + \left(p^2 - \frac{m^2}{\rho^2}\right)J(p\rho) = 0 \quad (1b)$$

Eq. (1a) is quantum well-type problem with the discontinuity of the effective mass at the interface taken into account, and Eq. (1b) is the differential equation of the Bessel functions. As result of the azimuthal symmetry, $m$ is an integer. The additional requirement of regularity imposes as solutions of Eq. (1b) the Bessel functions of first kind, $J_m$. The boundary condition $\psi(\rho = R) = 0$ imposes an additional restriction, namely, $p = \chi_m^l/R$, with $\chi_m^l$ the zero of $l$ order for $J_m$. The quantum well-type problem has: i) *even* solutions, $f(nz) = C\cos(k_w^n z)$ for $|z| \leq d$ and $A\exp[k_b^n(\mp z + d)]$ for $|z| > d$ with the boundary conditions $C\cos(k_w^n d) = A$, $Ck_w^n \sin(k_w^n d)/m_*^w = Ak_b^n/m_*^b$, and consequently, $\tan(k_w^n d) = m_*^w k_b^n/(m_*^b k_w^n)$, and ii) *odd* solutions $f(nz) = D\sin(k_w^n z)$ for $|z| \leq d$, and $\pm B\exp[k_b^n(\mp z + d)]$ for $|z| > d$ with the boundary conditions $D\sin(k_w^n d) = B$, $Dk_w^n \cos(k_w^n d) = -Bk_b^n$, and consequently $\cot(k_w^n d) = -m_*^w k_b^n/(m_*^b k_w^n)$; the sign $-$ and $+$ corresponds to positive and negative values of $z$, respectively (the origin is the cylinder center), and $k_b^n = \sqrt{2m_*^b[V_b - \varepsilon(n)]/\hbar^2}$, $k_w^n = \sqrt{2m_*^w \varepsilon(n)/\hbar^2}$, with $m_*^w$, $m_*^b$ the effective masses of carriers inside (well), outside (barrier) of QD. Energies $\varepsilon(n)$ are obtained numerically from the boundary conditions which have been written to express continuity of the wave function and the current-conserving at the interface. $n$ is the index for the possible states in $z$ direction.

For the bound states solutions of Eqs. (1) in the embedded material we consider those respecting restriction $V_b^* \geq E$. Then the energy levels corresponding to a given $n$ are found by applying the assumed parabolic dispersion law to the inequality

$$V_b^* \geq E_{nml} = \varepsilon(n) + \frac{1}{2m_*}\left(\frac{\hbar \chi_m^l}{R}\right)^2 \quad (2)$$

Consequently, according to the three quantum numbers introduced to characterize the single particle state, *nml*, the single particle envelope wavefunctions have the form

$$\psi_{nml}(\mathbf{r}) = C_{nml} J_m\left(\frac{\chi_m^l}{R}\rho\right) f(nz) e^{im\varphi} \quad (3a)$$

where the normalization constant is

$$C_{nml} = \sqrt{\frac{1}{2\pi}\frac{2R^{-2}}{[J_{m+1}(\chi_m^l)]^2}} \quad (3b)$$
$$\times \begin{cases} C \text{ (even) or } D \text{ (odd), for } |z| < d \\ A \text{ (even) or } B \text{ (odd), for } |z| \geq d \end{cases}$$

The different effective masses and confinement potentials in $z$ direction of the carriers make their corresponding envelope wavefunctions different. As we already mentioned, the QD model adopted here neglects the Coulomb interaction, an approximation which holds for small QDs (strong geometrical confinement), where the inter-level energy of the electron and hole is larger than the Coulombic interaction [12]. For the EHP states that are obtained after light irradiation, we consider the pairs which are correlated by the *optical* selection rules. The EHP has the envelope wave vector given by

$$X_{\alpha\beta}(\mathbf{r}_e,\mathbf{r}_h) = \psi_\alpha(\mathbf{r}_e)\psi_\beta(\mathbf{r}_h), \qquad (4a)$$

and the energy,

$$E_{\alpha\beta} = E_\alpha + E_\beta + E_g, \qquad (4b)$$

where $\alpha, \beta$ is the set of quantum numbers $n_i m_i l_i$, with $i=e, h$, respectively. The optical selection rules for inter-band transitions are obtained via the electric dipole matrix element $\langle \Psi_{\alpha\sigma} | \boldsymbol{\varepsilon} \cdot \mathbf{r} | \Psi_{\beta\tau} \rangle$. Thus,

$$\begin{aligned}\langle \Psi_{\alpha\sigma} | \boldsymbol{\varepsilon} \cdot \mathbf{r} | \Psi_{\beta\tau} \rangle &= \Omega_0^{-1} \\ &\times \int_{\Omega_0} d\mathbf{r}\, u_\sigma^*(\mathbf{r})\, \boldsymbol{\varepsilon}\cdot\mathbf{r}\, u_\tau(\mathbf{r}) \int_V d\mathbf{r}\, \psi_\alpha^*(\mathbf{r})\psi_\beta(\mathbf{r}),\end{aligned} \qquad (5)$$

where $V$ is the volume of the system, and $\Omega_0$ is the volume of primitive cell. By re-introducing the index $e$ or $h$ to denote the type of carrier, the integral over the envelope wave functions from Eq. (5) yields

$$\begin{aligned}&\int_V d\mathbf{r}\, \psi_{nml}^*(\mathbf{r})\psi_{n'm'l'}(\mathbf{r}) \\ &= \int_0^R d\rho\, \rho\, J_m\!\left(\frac{\chi_m^l}{R}\rho\right) J_{m'}\!\left(\frac{\chi_{m'}^{l'}}{R}\rho\right) \\ &\quad \times \int_{-\infty}^{\infty} C_{ml}^e C_{m'l'}^h\, dz\, f^e(nz) f^h(n'z) \int_0^{2\pi} d\varphi\, e^{i(m-m')\varphi} \\ &= \delta_{mm'}\delta_{ll'} \int_{-\infty}^{\infty} C_{ml}^e C_{m'l'}^h\, dz\, f^e(nz) f^h(n'z)\end{aligned} \qquad (6)$$

where $n = n_e$, $m = m_e$, $l = l_e$, and $n' = n_h$, $m' = m_h$, $l' = l_h$. Eq. (6) establishes the transitions are allowed between electron and hole states with $l = l'$, and $m = m'$. We consider the EHPs for the ground state energy in $z$ direction (see the below argument), that is, $n_e = n_h = 1$. Consequently, dismissing presence of the quantum number $n$ we re-denote the EHP states envelope wavefunction by $X_\alpha(\mathbf{r}_e,\mathbf{r}_h) = \psi_{ml}^e(\mathbf{r}_e)\psi_{ml}^h(\mathbf{r}_h)$, where $\alpha$ stands for the quantum numbers $ml$.

By introducing $X_\alpha^+ (X_\alpha)$, creation (annihilation) operators of EHP by relation $X_\alpha^+|0\rangle = |X_\alpha\rangle$, where $|0\rangle$ is the vacuum state (no EHP), the EHP-phonon coupling reads [13]

$$H_{X-LO} = \sum_{m,l;\, \alpha_1,\alpha_2} M_{ml}^{\alpha_1\alpha_2} X_{\alpha_1}^+ X_{\alpha_2} \left(a_{ml} + a_{ml}^+\right) \qquad (7)$$

with $M^{\alpha_1\alpha_2}$ the Hermitian EHP-phonon operator coupling the EHP states, $X_{\alpha_1}$ and $X_{\alpha_1}$. For the adopted adiabatic limit, which assumes no mixing of the carrier states by LO phonons (see Appendix A), we have $\alpha_1 = \alpha_2$.

The Huang-Rhys factor for confined phonon modes is defined by [14]

$$g^\alpha = \sum_{ml} g_{ml}^\alpha = \sum_{ml} \frac{|M_{ml}^{\alpha\alpha}|^2}{\hbar^2 \omega_{LO}^2} \qquad (8)$$

Regarding relation with the experiment, according to the theory of optical spectra (see, e.g. [3]), the Huang-Rhys factor introduced by Eq. (8) is expected to be obtained in the experiment as the ratio of intensities of 1LO and 0LO lines only if the optical transitions are adiabatic.

### 3. Confined optical phonon modes

To describe the electron-LO phonon coupling in nanostructures, following the works of Licari and Evrard [15], Mori and Ando [16], Huang and Zhu [17] or Klein et al. [18], we consider dielectric continuum model and Fröhlich-type interaction between carriers and the LO phonons of a polar semiconductor. The following description addresses type I semiconductor QDs, where both photo-excited carriers, the electrons and holes are confined in the QD volume.

The necessary relations in the following discussion within the dielectric continuum model are as follows [18]:

$$\frac{1}{\varepsilon_\infty} - \frac{1}{\varepsilon_0} = \frac{\omega_P^2}{(1+2\beta)^2 \omega_{LO}^2}, \qquad (9a)$$

and

$$\nabla\phi = \frac{4\pi n\mathrm{e}}{(1+n\chi 8\pi/3)}\mathbf{u}, \qquad (9b)$$

where $\varepsilon_0$ is the static dielectric constant, $\varepsilon_\infty$ is the high frequency dielectric constant, $\omega_{LO}$ is the LO phonon frequency, $\omega_P^2 = 4\pi n\mathrm{e}^2/\mu$ is the ion plasma frequency with $\mu$ the ion reduced mass, $\beta = 4\pi n\chi/3$ with $\chi$ the polarizability of an ion pair and $n$ the volume density of ion pairs, $\mathbf{u} = \mathbf{u}_+ - \mathbf{u}_-$ is the relative displacement of the positive and negative ions. $\phi$ is the polarization potential derived from the macroscopic electric field, $\mathbf{E}$. $\mathbf{D}$ is the electric displacement. For the reader's convenience, derivation of Eqs. (9) is presented in Appendix B.

Next, following the works dealing with solution for the polarization potential induced by the electron-LO phonon interaction for cylindrical shapes of QDs (see Refs. [7, 8]), we solve $\nabla \cdot \mathbf{D} = \varepsilon \nabla \cdot \mathbf{E} = 0$, assuming the dielectric constant $\varepsilon$ is space independent. Thus, one obtains $\varepsilon \Delta \phi = 0$. This equation has two types of solutions, namely, bulk-like modes solution, when $\varepsilon = 0$, and surface modes solution, when $\Delta \phi = 0$. Since for spherical QDs the surface modes do not contribute to the electron-phonon interaction [19], we expect a less important contribution of these modes to the exciton-phonon interaction when the aspect-ratio, $R/d$, in the cylindrical QDs is not much larger than the unit. In the adiabatic approach the electron is fast oscillating and it experiences in the case of such shapes with not large aspect-ratio a negligible (in a first approximation) average surface ionic polarization charge. This results in a weak exciton-phonon interaction. In addition, the mechanical factor which is considered in the multimode dielectric continuum models (see, e.g., Refs. [10, 20]), the strain is of less importance for GaAs/AlAs, which is an almost lattice-matched heterostructure. Consequently, in what follows we consider in estimation of the Huang-Rhys factor only the confined bulk-like modes.

To describe the confined bulk-like LO phonon modes, we take into account the case of the photo-excited carriers confined in the embedded QD material. That is, by means of Eq. (B5), we search solution for

$$\varepsilon_1(\omega_{1LO}) \Delta \phi_1(\mathbf{r}) = 0, \quad \varepsilon_1(\omega_{1LO}) = 0 \quad (10a)$$

$$\varepsilon_2(\omega_{1LO}) \Delta \phi_2(\mathbf{r}) = 0, \quad \varepsilon_2(\omega_{1LO}) \neq 0 \quad (10b)$$

where $\varepsilon_1$, $\varepsilon_2$ stand for the embedded and embedding material, respectively. A solution of Laplace equation in cylindrical coordinate is of the form

$$\sum_{m=0}^{\infty} [A_m J_m(k\rho) + B_m N_m(k\rho)] (C_m e^{kz} + D_m e^{-kz}) e^{im\varphi}$$

where $N_m$ is the Neumann function. To have cancellation of the potential at $z \to \pm \infty$, the external potential, $\phi_2(\mathbf{r})$, which according to Eqs. (10b) is solution of the Laplace equation should be of the form

$$\phi_2(\mathbf{r}) = \begin{cases} \sum_m [A_m^+ J_m(k\rho) + B_m^+ N_m(k\rho)] D_m^+ e^{-kz} e^{im\varphi}, \text{if } z > d \\ \sum_m [A_m J_m(k\rho) + B_m N_m(k\rho)] (C_m e^{kz} + D_m e^{-kz}) e^{im\varphi}, \\ \text{if } -d \leq z \leq d \\ \sum_m [A_m^- J_m(k\rho) + B_m^- N_m(k\rho)] C_m^- e^{-kz} e^{im\varphi}, \text{if } z < -d \end{cases}$$

(10c)

The boundary conditions are as follows.
$$\phi_1(R, z, \varphi) = \phi_2(R, z, \varphi), \quad \phi_1(r, \pm d, \varphi) = \phi_2(r, \pm d, \varphi), \quad (10d)$$

$$\varepsilon_1 \frac{\partial \phi_1}{\partial \rho}\bigg|_R = \varepsilon_2 \frac{\partial \phi_2}{\partial \rho}\bigg|_R \Rightarrow \frac{\partial \phi_2}{\partial r}\bigg|_R = 0$$

$$\varepsilon_1 \frac{\partial \phi_1}{\partial z}\bigg|_{\pm d} = \varepsilon_2 \frac{\partial \phi_2}{\partial z}\bigg|_{\pm d} \Rightarrow \frac{\partial \phi_2}{\partial z}\bigg|_{\pm d} = 0 \quad (10e)$$

Eqs. (10d, e) impose $C_m^- = D_m^+ = C_m = D_m = 0$, that is, $\phi_2(\mathbf{r}) = 0$. $\phi_1$ may be any function, which respects boundary conditions and is finite. Then, by considering a similar expansion as that for $\phi_2$, that is, $\phi_1(\mathbf{r}) = \sum_{ml} \phi_{1ml}(\mathbf{r})$, the boundary conditions from Eq. (10e) impose a solution with two quantum numbers, $m$, $l$, for the polarization potential of the QD material,

$$\phi_1(\mathbf{r}) = \sum_{ml} \phi_{1ml}(\mathbf{r})$$

$$= \sum_{ml} A_{1ml} J_m\left(\frac{\chi_m^l \rho}{R}\right) \begin{cases} \cos\left(\frac{l\pi}{2d}z\right) e^{im\varphi}, \text{if } l = 1, 3, 5... \\ \sin\left(\frac{l\pi}{2d}z\right) e^{im\varphi}, \text{if } l = 2, 4, 6... \end{cases}$$

(11)

Comparatively to Ref. [7] in Eq. (11) the radial function enters as linear combinations of $J_m(\chi_m^l \rho/R)$ instead of $J_0(\chi_0^l \rho/R)$. The quantum mechanical form of Eq. (11) is obtained by expressing the classical relative displacement as Hermitian quantum field operators [21]. Thus, by introducing the creation (annihilation) $a_{ml}^+$ ($a_{ml}$) operators of the $ml$ oscillator mode one obtains for $r \leq R$ (see Appendix C)

$$\phi_{1ml}^{s}(\mathbf{r}) = \sqrt{\left(\frac{1}{\varepsilon_\infty} - \frac{1}{\varepsilon_0}\right)}$$

$$\times \sqrt{\frac{8\hbar\omega_{LO}}{J_{m+1}^2(\chi_m^l)(\chi_m^l)^2 d\left[1 + (\pi dR)^2/\left(2(d\chi_m^l)^2\right)\right]}}$$

$$\times \left[\lambda_{ml}^s(\mathbf{r})a_{ml}^s(\mathbf{r}) + h.c.\right]$$

$$\equiv f_{ml}\left[\lambda_{ml}^s(\mathbf{r})a_{ml}^s(\mathbf{r}) + \lambda_{ml}^{s*}(\mathbf{r})a_{ml}^{s+}(\mathbf{r})\right] \quad (12)$$

where super-script '$s$' holds for *odd* or *even* $l$.

**4. Exciton-LO phonon interaction. Application**
From [13] we infer the explicit form of the EHP-phonon coupling as follows

$$M_{ml}^{\alpha_1\alpha_2} = -e f_{ml}$$
$$\times \int \langle 0|X_{\alpha_1}[\lambda_{ml}(\mathbf{r}_e) - \lambda_{ml}(\mathbf{r}_h)]X_{\alpha_2}^+|0\rangle d\mathbf{r}_e d\mathbf{r}_h ,$$
$$= -e f_{ml}\delta_{l_1l_2}\delta_{m_1m_2}\left(K_{ml;m_1l_1;m_2l_2}^e - K_{ml;m_1l_1;m_2l_2}^h\right) \quad (13a)$$

with

$$K_{ml;\alpha_1;\alpha_2}^e = \int \psi_{m_1l_1}^{e*}(\mathbf{r}_e)\psi_{m_2l_2}^e(\mathbf{r}_e)\lambda_{ml}(\mathbf{r}_e)d\mathbf{r}_e , \quad (13b)$$

where $\alpha_i = m_i l_i$, and $ml$ are the couple of quantum numbers describing the EHP state, and the electron-phonon interaction potential, respectively; similar relation holds for holes. The Kronecker symbols in Eq. (13a) result from integration of the orthonormal electron or hole single particle states. From Eqs. (13), after integration over the azimuthal angle $\varphi$, we obtain

$$K_{ml;\alpha_1;\alpha_2}^e = 2\pi\delta_{0,m}\int_0^R d\rho\rho J_m\left(\frac{\chi_m^l\rho}{R}\right)\left[J_{m_1}\left(\frac{\chi_{m_1}^{l_1}\rho}{R}\right)\right]^2$$

$$\times \begin{cases} \int_{-d}^{d} dz \left|C_{m_1l_1}^e\right|^2 \cos\left(\frac{l\pi}{2d}z\right)[f^e(z)]^2, \\ \text{for } l \text{ even} \\ \int_{-d}^{d} dz \left|C_{m_1l_1}^e\right|^2 \sin\left(\frac{l\pi}{2d}z\right)[f^e(z)]^2, \\ \text{for } l \text{ odd} \end{cases}$$

(14)

and only the terms with $m = 0$ of the LO phonon modes are involved in the EHP-phonon coupling. Integration in Eq. (14) is done over the QD volume, where the electric potential is non-zero.

As an application of the theory, we model GaAs/AlAs QD (type I heterostructure), for which we consider the following numerical values [22]: a) for GaAs the isotropic effective masses are $m_e^{\text{GaAs}} = 0.067m_0$, $m_h^{\text{GaAs}} = 0.45m_0$, the bulk energy gap is $E_g = 1.42\text{eV}$, $\varepsilon_0 = 12.8$, $\varepsilon_\infty = 10.1$; b) for AlAs, the isotropic effective masses are $m_e^{\text{AlAs}} = 0.146m_0$, $m_h^{\text{AlAs}} = 0.76m_0$; c) the band-offsets are $V_b^e = 0.75\text{eV}$, $V_b^h = 0.6\text{eV}$; d) the energy of the LO phonon of GaAs is $\hbar\omega_{LO} = 0.036\text{eV}$. We consider the values of QD height $d = 2.6\text{nm}$ and radius $R = 3 \div 8\text{nm}$. For the radius range we consider, we obtain one electron level (even wave function) and three hole levels (the closest of the bottom is an even wave function) for the quantum well problem (Eq. 1a). Assuming statistical population of the quantum well energy levels, we consider only the first electron and hole states are occupied. As GaAs/AlAs heterostructure is a lattice-matched heterostructure the fabricated QD [23] is an almost strain-free nanostructure, and to calculate the exciton ground state energy, we consider for the gap energy of GaAs its bulk value. In this case the predicted ground state exciton energy has values in the range reported in Ref. [23].

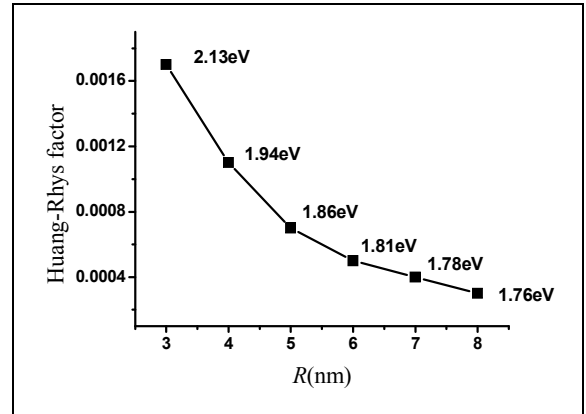

**Figure 1** Huang-Rhys factors as function of the cylinder radius of GaAs/AlAs QD with height $2d = 2.6\text{nm}$. The numbers show the predicted exciton ground state energy for the QDs of different radii $R$.

The resulting energies of the EHPs corresponding to the ground excitonic state and the Huang-Rhys factors are shown in Figure 1. The number of exciton states increases from one state for $R = 3\text{nm}$ to 7 states for $R = 8\text{nm}$. The calculus shows that the inter-level exciton energy established in the QD is at least 2.5 times larger than the LO phonon energy for the radius range we considered. Thus, calculus of the Huang-Rhys factor with Eq. (8) is fairly done within the adiabatic treatment we considered for the small QDs, in the range specified in Figure 1. The results show the Huang-Rhys factor is larger for smaller QDs, the typical behavior reported in the literature, see e.g. [3, 4, 24].

## 5. Conclusions

In the summation over the bulk-like LO phonon modes $ml$ from Eq. (8) practically only the mode 02 contributes, the others contributions being several orders of magnitude smaller. The magnitude of the Huang-Rhys factors predicted by the present QD model is included in the range 0.001-0.01 reported in Ref. [24] for such kind of QDs. In addition, the present treatment predicts existence of a representative LO phonon mode stronger coupled to the exciton in QD.

Extension of study of the representative phonon mode(s) to the case of exciton states mixed by LO phonons (non-adiabatic treatment) would be of interest. Dynamics of relaxation after light irradiation in QD can be described within a model where the acoustic phonons are a reservoir interacting with the exciton-LO phonon system having a reduced number of degrees of freedom.

## Appendix A

In this Appendix, following [25], we map the electron-lattice Hamiltonian from the configuration representation to the representation in the second quantization.

Then, we discuss some aspects regarding the adiabatic or non-adiabatic character of the optical transitions in QDs. For clarity of discussion, we consider an electronic system with the ground state $|g\rangle$ and two excited states $|i\rangle, |f\rangle$, in the diabatic representation, interacting with phonons. The total Hamiltonian reads

$$H = |g\rangle H_g \langle g| + |i\rangle H_i \langle i| + |f\rangle H_f \langle f| + H_{if}(|i\rangle\langle f| + |f\rangle\langle i|), \quad (A.1)$$

where $H_g$, $H_i$, $H_f$, and $H_{if}$ are the ground, excited states, and interaction between the excited states Hamiltonians. We consider for simplicity a single nuclear coordinate, the nuclear displacement relative to the equilibrium position, $Q$, having the origin at the minimum of the ground state. We denote the frequency of vibration by $\omega$. Assuming a $Q$ linear dependence of $H_i$, $H_f$, and $H_{if}$ and a parabolic shape of the potential energy surfaces (PESs), we write

$$H_g = P^2(2m)^{-1} + m\omega^2 Q^2/2,$$
$$H_{if} = \sqrt{2}(m\omega/\hbar)^{-1/2} C_{if} Q,$$
$$H_i = \varepsilon_i + P^2(2m)^{-1} + m\omega^2(Q+Q_i)^2/2,$$
$$H_f = \varepsilon_f + P^2(2m)^{-1} + m\omega^2(Q+Q_f)^2/2,$$

where $\varepsilon_i$ is the zeroth-order energy separation between the $i$-th excited PES and the ground state PES (similarly for $\varepsilon_f$), and $C_{if}$ is a constant. Note that $H_{if}$ is written in non-Condon approximation, that is, the nuclear coupling between the two excited states is $Q$ dependent. Next, we introduce the dimensionless momentum $p = (\hbar m\omega)^{-1/2} P$, and dimensionless coordinate $q = (m\omega/\hbar)^{1/2} Q$, and obtain

$$H_g = \frac{\hbar\omega}{2}(p^2 + q^2), \quad H_i = \varepsilon_i + \frac{\hbar\omega}{2}[p^2 + (q+q_i)^2],$$
$$H_f = \varepsilon_f + \frac{\hbar\omega}{2}[p^2 + (q+q_f)^2], \quad H_{if} = \sqrt{2} C_{if} q.$$

Then, we use the usual replacement $q = (a+a^+)/\sqrt{2}$, and $p = -i(a-a^+)/\sqrt{2}$, where $a$ ($a^+$) are the annihilation (creation) boson operators. Thus, one obtains

$$H_g = \hbar\omega(a^+ a + 1/2), \quad H_{if} = C_{if}(a^+ + a),$$
$$H_i = \varepsilon_i + \hbar\omega[(a^+ a + 1/2) + (a^+ + a)q_i/\sqrt{2} + q_i^2],$$
$$H_f = \varepsilon_f + \hbar\omega[(a^+ a + 1/2) + (a^+ + a)q_f/\sqrt{2} + q_f^2]$$

With the closure relation $|g\rangle\langle g| + |i\rangle\langle i| + |f\rangle\langle f| = \mathbf{1}$, one obtains

$$H = \hbar\omega a^+ a + |i\rangle\varepsilon_i'\langle i| + |f\rangle\varepsilon_f'\langle f|$$
$$+ (M_i |i\rangle\langle i| + M_f |f\rangle\langle f|)(a^+ + a), \quad (A.2)$$
$$+ C_{if}(|i\rangle\langle f| + |f\rangle\langle i|)(a^+ + a)$$

where $\varepsilon_i' = \varepsilon_i + \hbar\omega(q_i^2 + 1)/2$, and $M_i = \hbar\omega q_i/\sqrt{2} = \hbar\omega\sqrt{g_i}$, in which $g_i$ is the Huang-Rhys factor of the state $|i\rangle$ (similarly for the $|f\rangle$ state). Thus, in this single frequency model we obtain a simple dependence between the nuclear coupling and the Huang-Rhys factor, namely $M_i = \hbar\omega\sqrt{g_i}$. With introduction of the creation annihilation operators of electronic states by $|i\rangle\langle i| = C_i^+ C_i$ and $|i\rangle\langle f| = C_i^+ C_f$ (similarly for the $|f\rangle$ state), justified by $C_i^+ C_i |i\rangle = C_i^+ |g\rangle = |i\rangle$, $C_i^+ C_f |f\rangle = C_i^+ |g\rangle = |i\rangle = |i\rangle\langle f|f\rangle$, and $C_i^+ C_f |i\rangle = 0 = |i\rangle\langle f|i\rangle$, Eq. (A.2) reads

$$H = \hbar\omega a^+ a + \sum_{j=i,f} C_j^+ C_j \left[\varepsilon_j ' + M_j (a^+ + a)\right]$$
$$+ C_{if}\left(C_i^+ C_f + C_f^+ C_i\right)(a^+ + a) \quad (A.3)$$

Eq. (A.3) describes interaction between an electronic system and phonons. It is of the form of the localized defect model where the electronic states are mixed by phonons. The discrete structure of levels in this model is appropriate for describing the atomic-like energy structure of QDs, and this Hamiltonian is often adopted in QDs problems. Regarding the type of approach, an adiabatic treatment of QD implies absence in the Hamiltonian of the nuclear coupling terms between PESs, that is $C_{if} = 0$ or equivalently a non-mixing of the electronic states by phonons. It is worth noting that in the case of electron transfer reaction in chemical reaction, usually $H_{if}$ is taken $Q$-independent and the diabatic PESs are electronically coupled by the term of form $C_{if}(|i\rangle\langle f| + |f\rangle\langle i|)$.

**Appendix B**
Next, we survey the main relations related to the dielectric continuum model necessary in this work. Thus, we have Maxwell's equations for neutral systems

$$\mathbf{D} = \varepsilon\mathbf{E} = \mathbf{E} + 4\pi\mathbf{P}, \quad \mathbf{E} = -\nabla\phi, \quad \nabla\mathbf{D} = 0, \quad (B1)$$

Newton's equation for the relative motion of the positive and negative ions,

$$\mu\ddot{\mathbf{u}} = -\mu\omega_0^2 \mathbf{u} + e\mathbf{E}_{loc}, \quad (B2)$$

the polarization field equation

$$\mathbf{P} = ne\mathbf{u} + n\chi\mathbf{E}_{loc}, \quad (B3)$$

Lorentz's relation for the local field

$$\mathbf{E}_{loc} = \mathbf{E} + 4\pi\mathbf{P}/3, \quad (B4)$$

the frequency dependence of the dielectric constant [17],

$$\varepsilon(\omega) = \varepsilon_\infty \frac{\omega^2 - \omega_{LO}^2}{\omega^2 - \omega_{TO}^2}, \quad (B5)$$

and the Lyddane-Sachs-Teller relation

$$\frac{\varepsilon_0}{\varepsilon_\infty} = \frac{\omega_{LO}^2}{\omega_{TO}^2}, \quad (B6)$$

where the new introduced notations are as follows: $\mathbf{P}$, is the polarization corresponding to the macroscopic field $\mathbf{E}$; $\omega_0$ is the characteristic frequency of the ion pair, $\mathbf{E}_{loc}$ is the local field experienced by the ion pair; $\omega_{TO}$ is the transversal optical phonon frequency. With Eqs. (B) by considering irrotational and solenoidal solutions the frequency of the longitudinal and transverse oscillations, respectively, are obtained, that is, $\omega_{LO}^2 = \omega_0^2 + 2\omega_P^2/[3(1+2\beta)]$, and $\omega_{TO}^2 = \omega_0^2 - \omega_P^2/[3(1-\beta)]$. Considering the static case for Eqs. (B1-4), that is, setting $\mathbf{E} = 0$ and $(\varepsilon_0 - 1)\mathbf{E} = 4\pi\mathbf{P}$, and using Eqs. (B6) one obtains Eq. (9a). With Eqs. (B), for LO phonon modes ($\varepsilon = 0$ in Eqs. (B1, 3)) one obtains Eq. (9b).

**Appendix C**
By choosing

$$\mathbf{u}_{ml}^s(\mathbf{r}) = u_{ml}^0 \left[\nabla\left[\lambda_{ml}^s(\mathbf{r})\right]\right]$$

$$= \begin{cases} u_{ml}^0 \nabla\left[J_m\left(\frac{\chi_m^l \rho}{R}\right)\cos\left(\frac{l\pi}{2d}z\right)e^{im\varphi}\right] \\ = u_{ml}^0 \nabla\left[\lambda_{ml}^{odd}(\mathbf{r})\right] \text{ if } l = 1, 3, 5... \\ u_{ml}^0 \nabla\left[J_m\left(\frac{\chi_m^l \rho}{R}\right)\sin\left(\frac{l\pi}{2d}z\right)e^{im\varphi}\right] \\ = u_{ml}^0 \nabla\left[\lambda_{ml}^{even}(\mathbf{r})\right] \text{ if } l = 2, 4, 6... \end{cases} \quad (C1)$$

the condition $\phi_1(R) = 0$ is satisfied through Eq. (9b); the super-script '$s$' is added to differentiate between the *odd* and *even* cases. The orthogonality of the displacement *ml* modes reads

$$\int d\mathbf{r}\, \mathbf{u}_{m'l'}^s(\mathbf{r})\, \mathbf{u}_{ml}^{s*}(\mathbf{r})$$
$$= u_{m'l'}^0 u_{ml}^{0*} \int_V d\mathbf{r}\, \nabla\left[\lambda_{m'l'}^s(\mathbf{r})\right]\nabla\left[\lambda_{ml}^{s*}(\mathbf{r})\right] \quad . \quad (C2)$$
$$= |u_{ml}^0|^2 \pi\delta_{mm'}\delta_{ll'} \frac{J_{m+1}^2(\chi_m^l)}{4} \frac{2(d\chi_m^l)^2 + (\pi lR)^2}{d^2}$$

The above result of integration is found by writing the gradient operator in cylindrical coordinates and integrating over the volume of the cylinder.

Next, we determine the value of $|u_{ml}^0|^2$. The energy of LO phonon oscillators has the classical expression

$$H_{LO} = \frac{\mu}{2}\int d\mathbf{r}\, n\left(\mathbf{u}\mathbf{u}^*\omega_{LO}^2 + \dot{\mathbf{u}}\dot{\mathbf{u}}^*\right). \quad (C3)$$

The quantum mechanics version of Eq. (C3) reads

$$H_{LO} = \hbar\omega_{LO} \sum_{m,l} \left( a_{ml}^+ a_{ml} + \frac{1}{2} \right), \qquad (C4)$$

where the creation (annihilation) $a_{ml}^+$ ($a_{ml}$) operators of the $ml$ mode have been introduced. The passing to quantum mechanics of Eq. (C3) is done by introducing the relative displacement as Hermitian quantum field operators. Consequently,

$$H_{LO} = \frac{\mu}{2} \int d\mathbf{r}\, n \left[ \mathbf{u}(\mathbf{r})\mathbf{u}^+(\mathbf{r})\omega_{LO}^2 + \dot{\mathbf{u}}(\mathbf{r})\dot{\mathbf{u}}^+(\mathbf{r}) \right]. \quad (C5)$$

By using the derivative of field operators of the $ml$ mode

$$\dot{\mathbf{u}}_{ml}(\mathbf{r}) = i\left[ H_{LO}, \mathbf{u}_{ml}(\mathbf{r}) \right]/\hbar$$
$$= -i\omega_{LO} \sum_{ml} \left[ a_{ml}^+ \mathbf{u}_{ml}^*(\mathbf{r}) - a_{ml} \mathbf{u}_{ml}(\mathbf{r}) \right]$$

and its adjoint

$$\dot{\mathbf{u}}_{ml}^+(\mathbf{r}) = i\left[ H_{LO}, \mathbf{u}_{ml}(\mathbf{r}) \right]/\hbar$$
$$= i\omega_{LO} \sum_{ml} \left[ a_{ml} \mathbf{u}_{ml}(\mathbf{r}) - a_{ml}^+ \mathbf{u}_{ml}^*(\mathbf{r}) \right],$$

Eq. (C5) reads

$$H_{LO} = \frac{\mu}{2} n \omega_{LO}^2$$
$$\sum_{m,l,m',l'} \int d\mathbf{r}\, \left\{ \left[ a_{m'l'} \mathbf{u}_{m'l'}(\mathbf{r}) + a_{m'l'}^+ \mathbf{u}_{m'l'}^*(\mathbf{r}) \right] \right.$$
$$\times \left[ a_{ml} \mathbf{u}_{ml}(\mathbf{r}) + a_{ml}^+ \mathbf{u}_{ml}^*(\mathbf{r}) \right]$$
$$+ \left[ a_{m'l'}^+ \mathbf{u}_{m'l'}^*(\mathbf{r}) - a_{m'l'} \mathbf{u}_{m'l'}(\mathbf{r}) \right] \qquad (C6)$$
$$\left. \times \left[ a_{ml} \mathbf{u}_{ml}(\mathbf{r}) - a_{ml}^+ \mathbf{u}_{ml}^*(\mathbf{r}) \right] \right\}$$
$$= \frac{\mu}{2} n \omega_{LO}^2 \sum_{m,l,m',l'} \delta_{mm'} \delta_{ll'}$$
$$\times \int d\mathbf{r}\, \mathbf{u}_{ml}(\mathbf{r}) \mathbf{u}_{ml}^*(\mathbf{r}) \left( 2 a_{ml}^+ a_{ml} + 1 \right)$$

By comparing Eqs. (C4), (C5) and using Eq. (C2) one obtains

$$\left| u_{ml}^0 \right|^2 = \frac{4 d^2 \hbar}{\pi \mu n \omega_{LO} J_{m+1}^2(\chi_m^l) \left[ 2\left( d\chi_m^l \right)^2 + (\pi dR)^2 \right]} \quad (C7)$$

Next, with Eq. (9a, b) written for the oscillation mode $ml$, one obtains Eq. (12).

* Corresponding author: Tiberius O. Cheche
  E-mail: cheche@gate.sinica.edu.tw